\documentclass{article}
\usepackage{graphicx} 
\usepackage{style/subfigure} 
\usepackage{natbib}
\usepackage{algorithm}
\usepackage{algorithmic}
\usepackage{hyperref}
\usepackage{amsmath}

\usepackage[accepted]{style/icml2013}
\icmltitlerunning{Applying the Residue Number System to Neural Network Inference}
\newcommand{\Mod}[1]{\ \mathrm{mod}\ #1}
\begin{document} 

\twocolumn[
\icmltitle{Applying the Residue Number System to Network Inference}
\icmlauthor{Mohamed Abdelhamid}{mrhamid@mit.edu}
\icmlauthor{Skanda Koppula}{skoppula@mit.edu}

\vskip 0.2in
]


\section{Introduction} 
 
Use of neural networks for computer vision, speech recognition, and other applications has exploded in recent years, in part due to their unprecedented performance on a variety of benchmarks. Nonetheless, high-throughput and energy-efficient evaluation of such neural networks, and in particular, convolutional neural networks (CNNs), remains an active field of research. Evaluation of networks is memory and compute intensive, with the bottleneck depending on the network topology and layer types (convolutional or fully-connected [FCN]).

Significant effort has been made to reduce the memory footprint of neural networks, motivated by the fact that many off-chip memory accesses can dominate energy consumption during evaluation \cite{dorefa, prune, bnn}. This work explores the lesser studied objective of optimizing the multiply-and-accumulates executed during evaluation of the network. In particular, we propose using the Residue Number System (RNS) as the internal number representation across all layer evaluations, allowing us to explore usage of the more power-efficient RNS multipliers and adders. We motivate our optimization with Table 1, which summarizes the large number of multiply-and-accumulates (MACs) required during evaluation of popular networks. Small improvements to the underlying efficiency of the core multiply and accumulate block can have large improvements to the overall network evaluation.
\vspace{-2mm}
\begin{table}[h]
\caption{Computation Accounting for Popular Networks}
\label{sample-table}
\begin{center}
\begin{small}
\begin{sc}
\begin{tabular}{lcccr}
\hline
Net & MACs ($10^6$) & Params ($10^6$) \\
\hline
AlexNet    & 720 & 60 \\
GoogLeNet  & 1550 & 6.8 \\
SqueezeNet & 1700 & 1.25 \\
VGG-16 & 15300 & 138 \\
\hline
\end{tabular}
\end{sc}
\end{small}
\end{center}
\vskip -0.1in
\end{table}

Prior work applying RNS for efficient computation has largely focused on cryptographic applications and general purpose ALUs. In the machine learning domain, various optimizations such as Winograd and FFT transforms have been proposed to speed up network inference. As far as we are aware, this is the first attempt at applying RNS to neural network inference.

In section \ref{preliminaries}, we describe RNS and the requisite operations and methodology for RNS-based inference. Sections \ref{compare}, \ref{generation}, \ref{multiply}, and \ref{power} details our implementation of core RNS hardware modules in Verilog, and simulation results with estimated power, area, and latency. In section \ref{accuracy}, we describe accuracy estimates with our chosen RNS precision. In section \ref{break}, we tie together our results with an analysis of the break-even points at which it becomes beneficial to use RNS-based inference for low-power network evaluation. We conclude with a critique of our contributions.

All our Tensorflow, Bluespec/Verilog, and software model of digital hardware source code is available for re-use at \url{https://github.mit.edu/mrhamid/6888_Project}.

\section{Preliminaries}
\label{preliminaries}

\subsection{Residue Number System}
At its core, the Residue Number System relies on the Chinese Remainder Theorem (CRT) to represent a large integer as a tuple of smaller integers. This allows for more smaller and more parallelized arithmetic blocks. In RNS, an integer $X \Mod{M}$ with moduli set $\{m_1, m_2, \dots, m_k, \dots m_n\}$ is represented as $(x_1, \dots , x_n)$ where
$$
x_k \equiv X \Mod{m_k}
$$
The set of moduli is carefully chosen. In particular, the moduli are usually co-prime to reduce the number of distinct numbers that have the same RNS representation, and in our case, also chosen for hardware efficiency of operations. In this work, we fix our moduli to the structured 4-tuple $\{2^{n}\pm1, 2^{n+1} \pm 1\}$. This set can represent numbers in range $[0, M = \frac{(2^{2n}-1)(2^{2n+2}-1)}{3}]$ \cite{compare}. We choose the $n=7$ moduli set in this work. Every RNS number is stored using $7 + 8 + 8 + 9 = 32$ bits, and can fall in the range $[0, \frac{(2^{14}-1)(2^{16}-1)}{3} = 357886635]$. This is the representational range of a $28$-bit unsigned integer.

Addition of two RNS numbers $r^1$ and $r^2$ occurs element-wise: $r^s = r^1_k + r^2_k \Mod{m_k}$. Multiplication is similarly elementwise: $r^p = r^1_k \times r^2_k \Mod{m_k}$. Conversion in and out of RNS is based on CRT Theorems I, II, and III. Unfortunately, with the requisite divisions and iterative algorithms, conversion requires significant arithmetic overhead, as discussed by \cite{converter}.

\subsection{Piecing Together RNS Inference}
At its core, to construct complete end-to-end inference in RNS we need to support two operations: multiply-and-accumulate and ReLU. Our choice of moduli allow us to build on prior work for both: \cite{addition} proposes digital architecture for multiplication and addition $\Mod{2^n \pm 1}$, and \cite{compare} demonstrates comparison of RNS numbers $\Mod{2^n \pm 1}$ in VHDL. We use a comparison module to implement the ReLU nonlinearity.

We assume a discrete output for our network -- e.g. 10-output image-recognition with CIFAR-10. This allows us to avoid the overhead of conversion out of RNS at the end of the network; instead, with our comparator, we compute a $max$ over final layer softmax values, returning the index with the maximum value. All RNS operations occur in the realm of positive integers with fixed modulus. We explore this constraint in Section~\ref{accuracy}.

\section{Comparison in RNS}
\label{compare}
Though RNS multiplication and addition are operationally intuitive, comparing two RNS numbers is non-trivial. We follow the procedure given by \cite{compare}. The crux of the algorithm is reducing comparison to parity (even/odd) checking. To compare two unsigned integers $A,B \Mod{M}$, we compute the difference $C=A-B$ which becomes one of two values:
\[
  C =
  \begin{cases}
                                   A-B & \text{if $A \geq B$} \\
                                   M+A-B & \text{if $A < B$}
  \end{cases}
\]
Because with our chosen moduli $M$ is odd, these two values have different parities. As such, we can compute a comparison given a formula for the $\Mod{2}$ parity $X_P$ of an RNS number $X = (x_1, x^{*}_1, x_2, x^{*}_2)$. Parity is calculated with the following set of equations:
\begin{equation*}
\begin{split}
X_1 = x^{*}_1 + (2^n + 1) \times (2^{n-1}(x_1-x_1^{*}) \Mod{2^n-1}) \\
X_2 = x^{*}_2 + (2^{n+1} + 1) \times (2^{n}(x_2-x_2^{*}) \Mod{2^{n+1}-1}) \\
X_P = LSB(X_2) \oplus LSB((X_1 - X_2) \Mod{2^{2n}-1})
\end{split}
\end{equation*}
Proof is provided by \cite{compare}.
This is the \textit{full-comparator} we implement in combinational logic to execute the final layer maximum. In the ReLU, we are able to trim this combinational logic, because we compare with a fixed threshold $\frac{M}{2}$ (0 in our modulus world). We call this trimmed comparator a \textit{half-comparator}. In particular, the parity of $B = \frac{M}{2}$ is fixed and pre-computed, as well as the value of the additive inverse $-B = -\frac{M}{2}$ we feed into the modulo adder.

\begin{figure}[h]
\begin{center}
\centerline{\includegraphics[width=0.8\columnwidth]{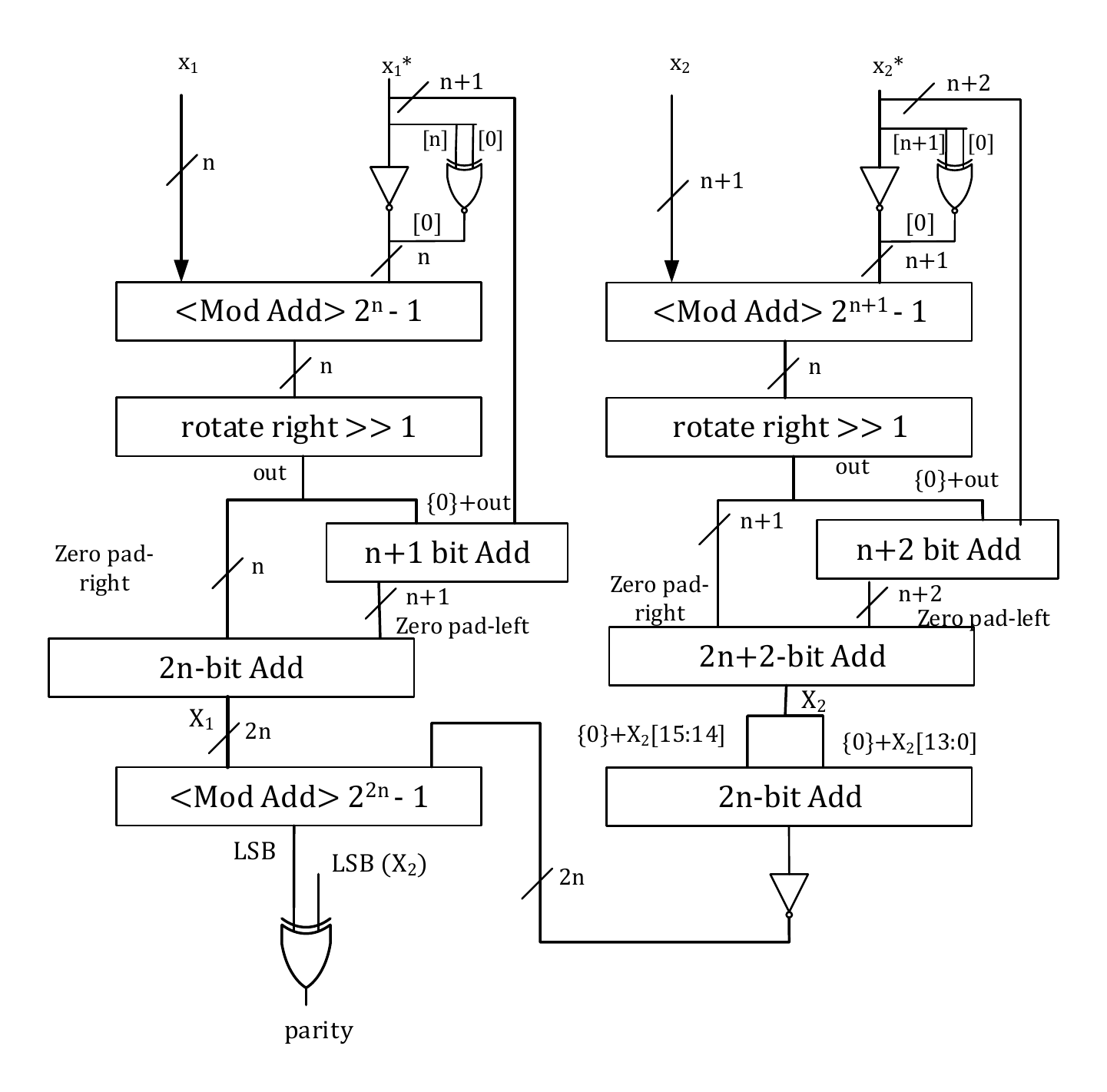}}
\caption{Combinational logic for calculating the parity of an RNS number $(x_1, x^{*}_1, x_2, x^{*}_2)$}
\label{fig:fullcompare}
\end{center}
\end{figure}

The parity-checking combinational logic implemented in Verilog is given by Figure~\ref{fig:fullcompare}. This is used in both the full and half comparator design. We modify the circuit given in \cite{compare}, which we suspect, from our testing, does not evaluate correctly in an exhaustive sweep of all $\sim$3 billion inputs.

We include various optimizations. For example, our choice of modulus allows for use of an inverter to find the additive inverse. It also allows for calculating the remainder with a 16 bit number as a single addition (bottom-right). Additionally, we implement multiplication with $2^n \Mod{2^{n+1}-1}$ as a right rotate. 
\vspace{-4mm}
\vspace{-12mm}
\section{Converting Input into RNS}
\label{generation}
For the chosen moduli set, the residue generation relies on the periodicity of the binary weights ($2^i$) in the modulus domain. As pointed out by~\cite{resgen}, the residue of a binary number is given by:
\begin{equation*}
    X \mod (2^{n_1}-1) = \sum_{i=0}^{n-1} 2^i x_i \mod (2^{n_1}-1)
\end{equation*}
taking the modulus causes the binary weights to repeat themselves when $i>n_1$. Therefore, the residue is calculated by periodically folding back the higher weights and adding them to the lower weights using a tree of modulo adders.

\section{RNS Arithmetic}
\label{multiply}
CNNs perform a large number of computations such as multiplication and accumulation throughout the convolutional as well as the fully connected layers. To fully exploit the advantages of transforming the network to the RNS system, efficient modulo arithmetic circuits have to be designed. This section presents efficient modulo multiplication and addition implementation using the same moduli set as the comparison, namely \{$2^{n_1}\pm 1,\ 2^{n_2}\pm1$\}.
\subsection{Modulo Addition}
The main advantage of RNS arithmetic is that each residue operates separately in parallel without any carry propagation from one residue to the other. Our conjugate moduli set requires modulo ($2^n -1$) addition/multiplication as well as ($2^n +1$).

First, the modulo ($2^n -1$) addition can be expressed as conventional n-bit addition if the sum is less than the modulus, while a correction is added if the sum overflows the modulus as follows:
\begin{equation*}
\resizebox{0.9 \columnwidth}{!}{
    $(A+B)\text{ mod }(2^n -1) = 
    \begin{cases}
        A+B-(2^n-1)  \\
        \text{\ \ }=A+B+1 &\text{if A+B} \geq 2^n -1, \\
        A+B & \text{otherwise}
    \end{cases}$}
\end{equation*}
Since, the output carry ($c_{out}$) of an n-bit adder can be used to detect the overflow condition which determines whether to increment the sum or not, then such carry can be fed back into the adder as proposed by \cite{addition} for ($2^n -1$) addition.
\begin{equation*}
\resizebox{0.9 \columnwidth}{!}{
    $ (A+B)\mod (2^n - 1) = (A+B+c_{out}) \mod 2^n $}
\end{equation*}

Second, a similar analysis can be done for the modulo ($2^n +1$) addition to show that 
\begin{equation*}
\resizebox{0.9 \columnwidth}{!}{
    $ (A+B+1)\mod (2^n + 1) = (A+B+\overline{c_{out}}) \mod 2^n $}
\end{equation*}
where diminished-1 numbers can be used for the inputs, or a correction circuit is added to the output to account for the extra '1'.

\begin{figure}[ht]
\begin{center}
\centerline{\includegraphics[width=\columnwidth, trim=100 80 240 90, clip]{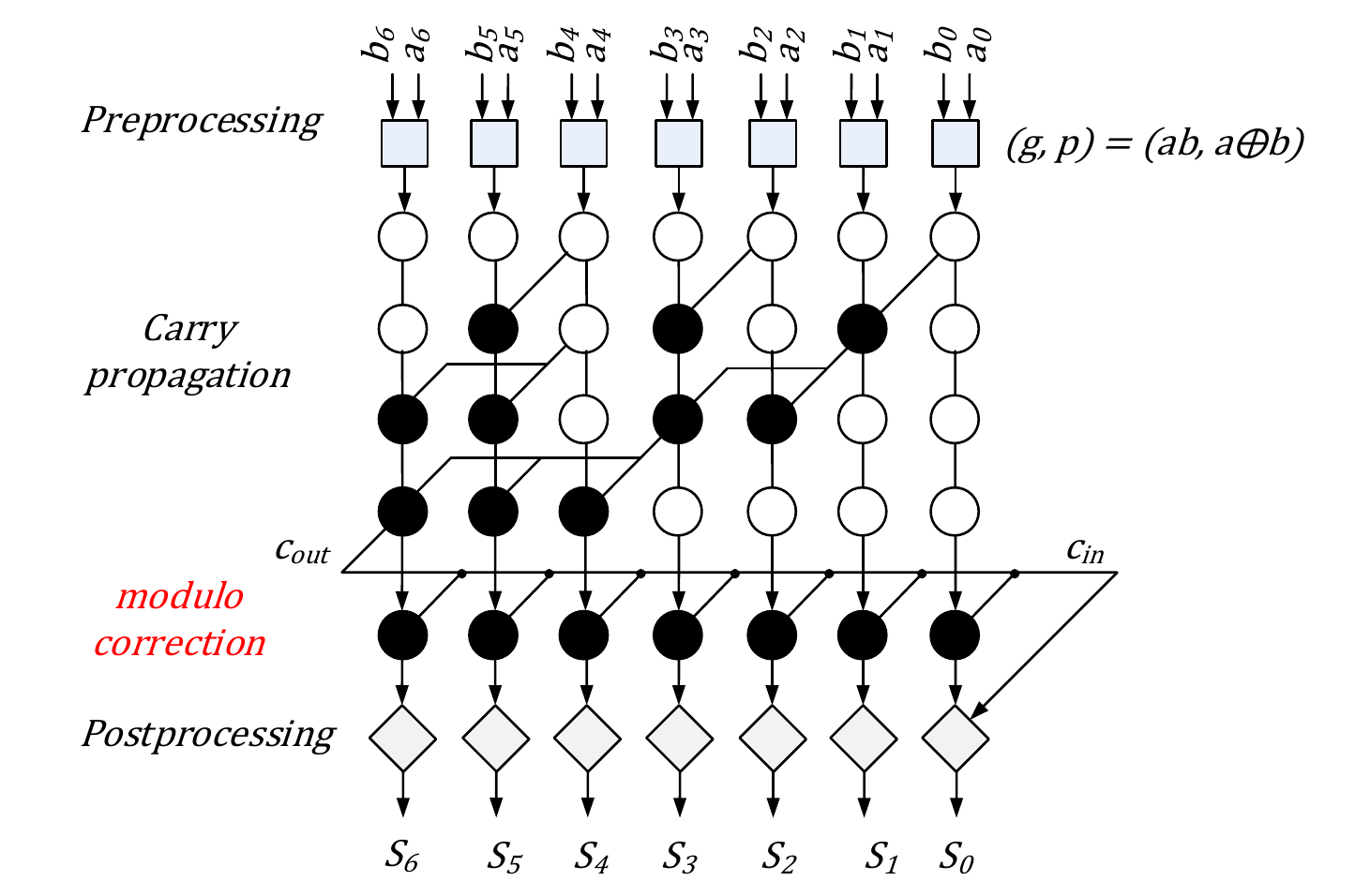}}
\caption{Parallel prefix adder for mod $2^7 - 1$ addition}
\label{fig:add}
\end{center}
\vskip -0.2in
\end{figure}

Since the addition in both moduli depends on the output carry, then fast parallel prefix adders are the most suitable implementation for the modulo adders. Figure~\ref{fig:add} shows the a modulo parallel prefix adder where the inputs are preprocessed into carry generate and propagate signals then a tree of fixed operation propagates the carry in only 4 levels. Each solid circle represents a dot operation which combines the group carry generate and propagate bits. Finally, a modulo end-around carry correction is required to feedback back the output carry ($c_{out}$) or its complement ($\overline{c_{out}}$) according to the designated modulus.
\vspace{-4mm}
\subsection{Modulo multiplication}
\begin{figure}[ht]
\begin{center}
\centerline{\includegraphics[width=0.9\columnwidth, trim=15 20 10 20, clip]{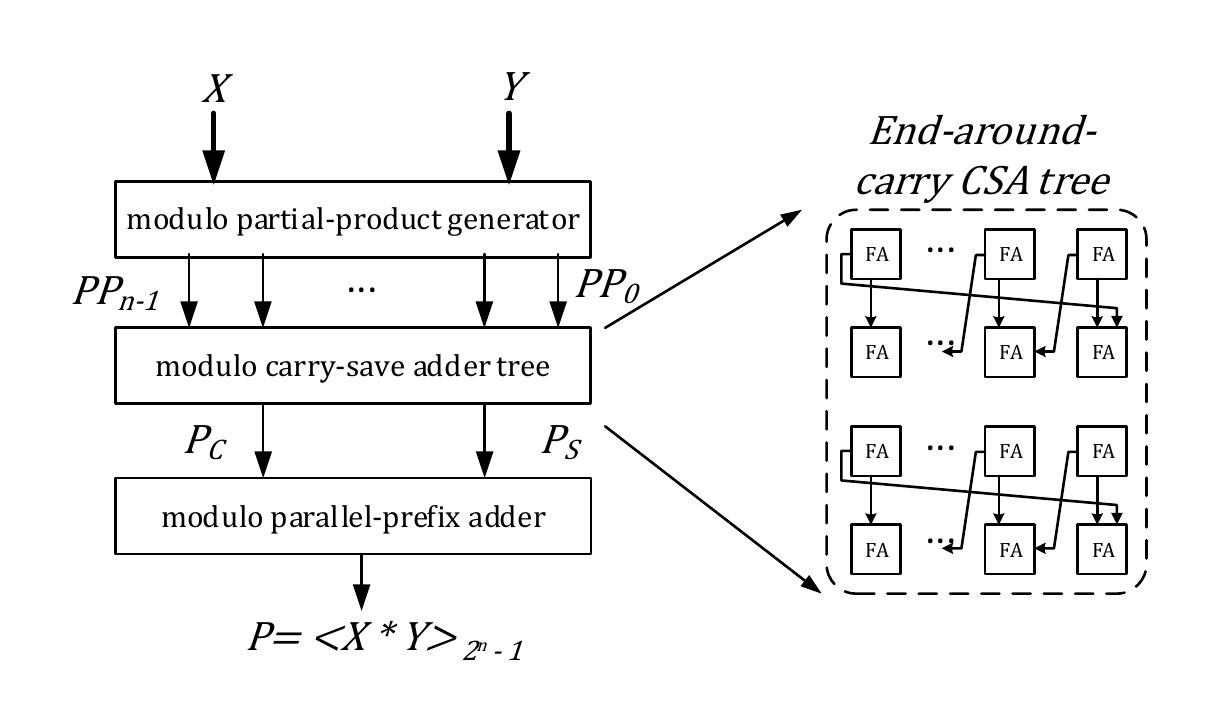}}
\caption{Mod $2^n - 1$ multiplier architecture showing the modulo carry save adder}
\label{fig:mult}
\vspace{-8mm}
\end{center}
\end{figure}

N-bit binary multiplication relies on generating N partial products and accumulating them all to produce the final product. Modulo multiplication relies on the same concept as well as the periodicity of the binary weights which causes the higher order partial products to rotate folding back into n-bit weights. Therefore, the partial products can be generated, similar to in~\cite{addition}, as
\vspace{-2mm}
\begin{equation*}
   X\cdot Y \mod (2^n-1) = \sum_{i=0}^{n-1} x_i \cdot (Y<<i) \mod (2^n-1)
\end{equation*}
\vspace{-1mm}
where the $<<$ operator represents a circular shift. Similarly, an expression for the partial products of the ($2^n+1$) modulus can be derived to be 
\begin{equation*}
    PP_i = x_i \cdot y_{n-i-1}\cdots y_0\overline{y_{n-1}}\cdots \overline{y_{n-i}} + \overline{x_i} \cdot 0\cdots01\cdots1
\end{equation*}
Such multipliers can be designed in a modular way where a block generates the required partial products according to the selected modulus. Then, a modulo carry save adder tree as shown in Figure~\ref{fig:mult} generates a redundant sum output ($P_C, P_S$) which is then added using a modulo adder to produce the final product.

\section{Results}
\subsection{RNS Power Consumption}
\label{power}
We built several building blocks for RNS-based network inference using Bluespec SystemVerilog and synthesized them using a commercial LP65nm CMOS process. Table~\ref{Syntable} shows the power and frequency of operation of the RNS blocks as well as their 32-bit counterparts. It is worth noting that the multiplier consumes almost half the power of the 32-bit block with a positive slack allowing for higher frequency of operation.

\vspace{-2mm}
\begin{table}[h]
\caption{Synthesis results}
\label{Syntable}
\begin{center}
\begin{small}
\begin{sc}
\begin{tabular}{lccccr}
\hline
Block & P ($mW$) & $f$ ($MHz$) & Slack($ps$) \\
\hline
Adder32    & 1.05 & 625 & 15.9    \\
AdderRNS  & 1.18 & 625 & 17.6     \\
Multiplier32 & 3.04 & 250 & 7.1   \\
MultiplierRNS & 1.56 & 250 & 95.4 \\
ConvertToRNS & 2.6 & 250 & 1.1     \\
ReLU-RNS & 0.88 & 156 & 109.5     \\
CompareRNS & 1.67 & 156 & 93.1    \\
\hline
\end{tabular}
\end{sc}
\end{small}
\end{center}
\vskip -0.1in
\end{table}

\subsection{Maintaining a Modulus Integer Network}
\label{accuracy}
A limitation using RNS is the necessity to maintain positive integer weights and activations within a given modulus $M$. To demonstrate the feasibility of this, and obtain a rough estimate of accuracy degradation when imposing such constraints, we train different flavors of a 8-layer (7 CNN/1 FC) network on the Street-View House Numbers (SVHN) dataset \cite{svhn}. We denote a (W, A)-FP/INT network as a network with W-bit weights and A-bit activations in either floating point or integer, respectively. Note that negative integers are interpreted as their respective positive value in a wrap-around modulus.

We first trained (32, 32)-FP. We used a set of shadow floating point weights, initialized to (32, 32)-FP, , and truncate these shadow weights in the forward pass to generate our (6, 6)-FP network (gradients get passed to the shadow weights). In our (32, 32)-INT and (6,6)-INT networks, we modify this truncation operation to be a suitable affine transformation to fit our bit width and desired range. Note that our activation function in the integer network changes to compare with $\frac{M}{2}$. Networks were trained for 15 epochs, with data augmentation, 50\% last layer dropout, and selecting the best model-checkpoint with highest validation accuracy. Note that a 6-bit integer is able to fit within each modulus of our RNS representation.

\vspace{-2mm}
\begin{table}[h]
\caption{SVHN Test Error Rate}
\label{acctable}
\begin{center}
\begin{small}
\begin{sc}
\begin{tabular}{lccccr}
\hline
Network & SVHN Test Error \\
\hline
(32, 32)-FP  & 3.95\%    \\
(6, 6)-FP &  6.69\%    \\
(32, 32)-Int & 4.54\% \\
(6, 6)-Int & 7.07\% \\
\hline
\end{tabular}
\end{sc}
\end{small}
\end{center}
\vskip -0.1in
\end{table}

As expected, reducing bitwidth increases error. Moving to integer networks appears to slightly decrease accuracy. The precise reason for this is unclear; perhaps, something wonky is occuring with the gradient updates and gradient magnitude. Networks were implemented in Tensorflow/Tensorpack.

\subsection{Estimation of the RNS Break-Even Point}
\label{break}
Use of RNS incurs overhead proportional to the output size, because of the comparatively expensive ReLU-RNS modules. If we have a $Y \times X$ fully-connected layer, we can compare the relative energy costs associated with performing the corresponding MACs and ReLU in RNS or non-RNS:
\begin{equation*}
\begin{split}
Y \times E_{RNSReLU} + XY \times (E_{RNSMult} + E_{RNSAdd}) \\ 
>  Y \times E_{ReLU} + XY \times (E_{Mult} + E_{Add})
\end{split}
\end{equation*}
$E_X$ is the energy per operation for hardware block $X$. Given our simulation results, this simplifies to:
\begin{equation*}
\begin{split}
X & > \frac{E_{ReLU} - E_{RNSReLU}}{(E_{RNSMult} + E_{RNSAdd}) - (E_{Mult} + E_{Add})} \\
  & \approx 0.98
\end{split}
\end{equation*}

This hints that it could be possible to achieve energy savings through RNS on FC layers of any size, because of our ReLU overhead/MAC savings ratio. It is demonstratable that the same result applies for a CNN layer, in which we would replace $X$ with $C_{in}K_XK_Y$, the size of channel-output filter. Note that this estimation is ignoring costs of memory accesses. Though, because of our choice of moduli, the amount of data being shuffled is similar in both systems.
\vspace{-2mm}
\section{Conclusion}
\label{conclusion}
In this work, we outlined use of the Residue Number System to perform inference on neural networks. Using our single-block implementation power estimates, we showed theoretical analysis of the advantages of RNS for an end-to-end system.
\subsection{Critique and Future Improvements}
In our next steps, we aim to demonstrate end-to-end inference stringing together our RNS blocks, comparing this system with a non-RNS system. It would be useful to explore other methods of translating networks to the integer domain, and testing accuracy drops in more networks and datasets. By fiddling with choice of adder design, we suspect that we can improve our RNS multipliers and comparator power.


\pagebreak

\bibliography{paper}
\bibliographystyle{style/icml2013}
\end{document}